\documentclass[%
 ,secnumarabic%
,amssymb, amsmath,nobibnotes, aps, prl]{revtex4}
\usepackage[dvipdfmx]{graphicx}
\usepackage{latexsym,mathrsfs}
\usepackage{dcolumn}
\usepackage{bm}

\def \av#1{\left\langle #1\right\rangle}

\renewcommand{\d}{{\rm d}}

\begin{document}

\title{Dynamics to the universal structure of one-dimensional self-gravitating systems in the quasi-equilibrium state}

\author{Tohru Tashiro}
\affiliation{Department of Physics, Ochanomizu University, 2-1-1 Ohtuka, Bunkyo, Tokyo 112-8610, Japan}

\date{\today}

\begin{abstract}
We investigate the quasi-equilibrium state of one-dimensional self-gravitating systems.
If the null virial condition is satisfied at initial time, it is found that the number density around the center of the system at the quasi-equilibrium state has the universality similar to two- and three-dimensional self-gravitating systems reported in \cite{Tashiro16,Tashiro10}. The reason why the null virial condition is sufficient for the universality is unveiled by the envelope equation. We present a phenomenological model to describe the universal structure by using a special Langevin equation with a
distinctive random noise to self-gravitating systems.
Additionally, we unveil a mechanism which decides the radius of the system.  
\end{abstract}

\maketitle

\section{Introduction}

Self-gravitating systems (SGSs) exhibit various interesting properties which systems with short-range forces do not have \cite{Levin14,Campa09}.
One of the examples is the presence of a stable state different from the equilibrium state.
In this Letter, we refer to the state as a quasi-equilibrium state (qES).
In general, the qES depends strongly on initial distributions.
If a null virial condition is fulfilled at initial time (i.e., all the velocities are zero), however, it has been found that the number densities of the two-dimensional SGS (2DSGS) or the three-dimensional SGS (3DSGS) at qES around the center of the system can be universally approximated by
\begin{equation}
\mbox{$\mathcal{N}$}(r) \simeq \frac{\mbox{$\mathcal{N}$}(0)}{(1+r^2/a^2)^\kappa}
\label{eq:fitf}
\end{equation}
with $\kappa \sim 1$ in 2DSGS \cite{Tashiro16} and $\kappa \sim 3/2$ in 3DSGS \cite{Tashiro10} regardless of the initial distribution in real space, where $r$ represents the distance from the center of the system.
The results of 2DSGS and 3DSGS coincide with the observations of  the number density of molecular clouds in  IC~5146 or the Taurus molecular complex \cite{Arzoumanian11,Stepnik03} and the number density of stars in globular clusters \cite{Binney08,Peterson75,Trager95}, respectively.
Thus, we can expect that if the null virial condition is satisfied, the value of $\kappa$ at qES is roughly the half of the dimension regardless of the initial distribution.
It is natural that the question arises as to whether the same universality holds in one-dimensional self-gravitating systems (1DSGSs).

1DSGS with $N$ particles is the simplest SGS whose two-body interaction force is constant independent of the distance between the particles, which will be demonstrated later.
The majority of studies about 1DSGS were aimed at investigating a relaxation to the thermal equilibrium state whose distribution function (DF) in phase space was obtained analytically in Ref.~\cite{Rybicki71}.
The relaxation time toward the state was originally estimated to be $N^2 t_d$ \cite{Hohl67,Hohl67b,Hohl68} where $t_d$ is the {\it dynamical time} which is the characteristic time for a particle to pass through the system.
After the studies, however, the relaxation time was found to depend on initial conditions.
If the initial distribution is a water-bag form, the relaxation time was revealed as much more than $N^2 t_d$ \cite{Wright82,Luwel85}.
On the other hand, for initial counterstreamed configuration with virial ratio 0.3, the system relaxes on the time scale $N^2 t_d$ \cite{Luwel84,Severne84,Reidl91}.
The relaxation process from the initial water-bag distribution was precisely described by the deviation from the equipartition law defined as
\begin{equation}
\Delta(t) = \frac{3}{5E}\sqrt{\frac{1}{N}\sum_{i=1}^{N}
\left[\overline{\mathcal{E}_i(t)}-\frac{5E}{3}\right]^2}
\label{eq:devi_equi}
\end{equation}
where $E$ is the total energy of the system and $\overline{\mathcal{E}_i(t)}$ is the time averaged one particle energy  per unit mass of $i$th particle until time $t$. It was unveiled that there are two different relaxation process, the microscopic relaxation occurring roughly after $Nt_d$ and the macroscopic relaxation occurring roughly after $4\times10^4Nt_d$ \cite{Tsuchiya94,Tsuchiya96}.

The other studies about 1DSGS dealt with the phase space distribution at qES. 
When the initial water-bag distribution is in virial equilibrium (i.e., the virial relation is one), the distribution at qES can be described well by the Lynden-Bell statistics which assumes ergodicity in phase space based on the Vlasov equation \cite{Lecar71,Yamaguchi08,Joyce11}
(Note that the Lynden-Bell statistics is not a unique theory for long-range interacting systems initially in the virial equilibrium. It has been reported that non-ergodic theories can describe the systems more accurately than the Lynden-Bell statistics \cite{deBuyl11,Ribeiro-Teixeira14,Benetti14}).
In contrast, if the initial distributions are far from virial equilibrium,  the distribution at qES cannot agree with  the statistics \cite{Lecar71,Yamashiro92,Yamaguchi08,Joyce11,Teles11}. 
The characteristic of the distributions from initial conditions near virial equilibrium is a {\it core-halo} structure.
The halo is composed of particles acquired high energy by a parametric resonance of the oscillating potential \cite{Teles11}.
Also, the distribution of particles losing energy approaches that of a degenerate Fermi gas, resulting in forming the core \cite{Teles11}.

Actually, in the previous researches about 1DSGS, the null virial condition was not focused on and the water-bag distributions were mainly adopted as the initial condition. 
In this Letter, therefore, we shall investigate qES of 1DSGS corresponding to the various initial distributions satisfying the null virial condition.
Although 1DSGS is simpler than 2DSGS and 3DSGS, it possesses the essential properties of SGSs.
Thus, if we could unveil the dynamics to qES of 1DSGS, by applying it to the other dimensional SGSs, 
we might clarify the origin of the universality of number density profiles of globular clusters or molecular clouds.

\section{One-dimensional self-gravitating system}

The one-dimensional gravitational potential per unit mass $\phi(x)$ generated by mass source $\rho(x)$ satisfies the following Poisson equation in the one-dimensional space. 
\begin{equation}
\frac{\d^2\phi(x)}{\d x^2}=4\pi{G}\rho(x)
\end{equation}
When the mass source is a mass point with mass $m$ at the origin, $\rho(x)=m\delta(x)$. 
Thus, a single-particle potential $\phi_{\rm sp}$ generated by the mass point satisfies the following Poisson equation,
\begin{equation}
\frac{\d^2\phi_{\rm sp}(x)}{\d x^2}=4\pi Gm\delta(x) \ ,
\end{equation}
and the potential can be solved as
\begin{equation}
\phi_{\rm sp}(x) = 2\pi{G}m|x| \ ,
\end{equation}
where we have used $\phi_{\rm sp}(x) = -\phi_{\rm sp}(-x)$ and $\phi_{\rm sp}(0)=0$.
Therefore, an interaction force between two mass points is constant, and then, the Hamiltonian of 1DSGS of $N$ particles with equal mass can be expressed as
\begin{equation}
H = \sum_{i=1}^{N}\frac{mu_{i}^2}{2} + 2\pi Gm^2\sum_{i>j}|x_i-x_j| \ ,
\end{equation}
where $x_i$ and $u_i$ mean a position and a velocity  of the $i$th particle, respectively.

\section{$N$-body simulations}

\subsection{Initial conditions}

As the initial distribution for $N$ body simulations, we shall adopt the following polytrope solution with a polytrope index $n$: 
\begin{equation}
f_n(r,v) \propto \left[\phi_n(\mathcal{R}_n) - \left\{\frac{1}{2}v^2+\phi_n(r)\right\}\right]^{n-1/2}\Theta(\mathcal{E}_n)
\end{equation}
where $r\equiv|x|$, $v\equiv|u|$, $\Theta$ is the Heaviside step function and  $\mathcal{R}_n$ means a radius of the system.
The potential energy per unit mass $\phi_n(r)$ satisfies the Poisson equation,
\begin{equation}
\frac{\d^2\phi_n(r)}{\d r^2}=4\pi Gm\mathcal{N}_n(r) \ ,
\end{equation}
where $\mathcal{N}_n(r) =\int\d vf_n(r,v) \propto \left\{\phi_n(R) -\phi_n(r)\right\}^n$.
Between the local pressure $p_n(r)=\int\d vv^2f_n(r,v)$ and the number density $\mathcal{N}_n(r)$, the polytrope relation, $p_n(r)\propto \mathcal{N}_n(r)^{1+1/n}$, is established like the three-dimensional polytrope solution \cite{Binney08}. 
Moreover, the total kinetic energy $K_n=\int{\d}r\int{\d v}\frac{1}{2}mv^2f_n(r,v)$ and the total potential energy $\Omega_n=\frac{1}{2}\int{\d}r\phi_n(r)m\mathcal{N}_n(r)$ satisfy the virial relation for 1DSGS, $2K_n/\Omega_n=1$.

The polytrope solution with $n=\infty$ corresponds to the thermal equilibrium state distribution.
Moreover, the number density of the the polytrope solution with $n=0$ in real space is constant, i.e., the water-bag distribution.
Hence, by changing the value of $n$, we can make various distributions which are physically characterized by the polytrope relation.

DF with the virial ration $V$ can be derived from $f_{n}(r,v)$ as $f_{n,V}(r,v) \equiv f_{n}(r,v/\sqrt{V})$ where the velocity interval of integration in phase space is changed to be $0\le v \le \sqrt{2V\left\{\phi_n(R) -\phi_n(r)\right\}}$. 
Note that the number density of $f_{n,V}(r,v)$ in real space, $\mathcal{N}_{n,V}(r)$, is equivalent to that of $f_{n}(r,v)$: $\mathcal{N}_{n,V}(r) = \mathcal{N}_{n}(r) \propto \left\{\phi_n(R) -\phi_n(r)\right\}^n$.

When we run the $N$-body simulations, we adopt a unit system where $4\pi G=M=1$ in which $M(=mN)$ is the total mass and the radius of the case with $n=0$ (water-bag distribution) is unity.

\subsection{Number densities at qES}

When the null virial condition is satisfied at initial time (i.e., the initial distribution in real space is $\mathcal{N}_{n}(r)$ and all the velocities are zero), the numerical simulations about 1DSGS make it clear that there are qESs and DF in phase space is symmetric about the origin.
Moreover, it is found that the number density in qES has the universality regardless of the initial polytrope index and the density around the center of the system can be fitted well by Eq.~(\ref{eq:fitf}).

\begin{figure}[h]
  \begin{center}
    \includegraphics[scale=0.851]{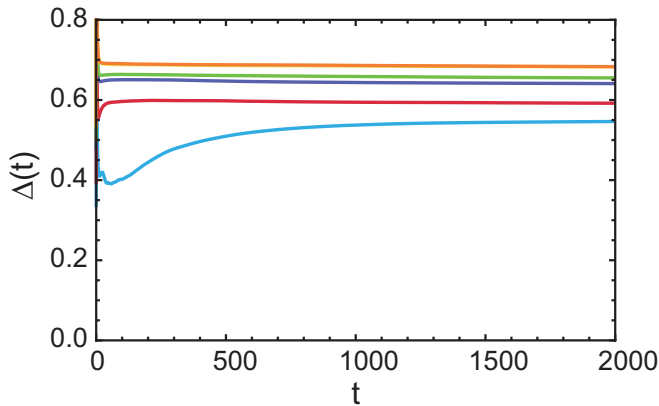}
    \caption{\label{fig:devi_equi}(color online) Time evolution of the deviation from the equipartition law defined by Eq.~(\ref{eq:devi_equi}). The five curves, from the lowest one to the highest one, correspond to $n=0$, 1, 5, 10 and 50 (100). The curve for $n=50$ overlay that for $n=100$ in the figure.}
  \end{center}
\end{figure}

We can make a decision whether the system is at qES by paying attention to
the change of physical quantities charactering the system in time, e.g. the total kinetic energy and the potential energy.
As one of the examples of such physical quantities, we show the time evolution of the deviation from the equipartition law, Eq.~(\ref{eq:devi_equi}),  for $n=0$, 1, 5, 10 and 50 (100) in Fig.~\ref{fig:devi_equi}.
The curve for $n=50$ overlay that for $n=100$ in the figure.
Although the behavior of the saturation of the deviation for $n=0$ (corresponding to the water-bag distribution) is exceptional, all the deviations does not change after a time and we regard this circumstance as qES.
Note that the behavior of $\Delta$  is completely different from that in Refs.~\cite{Tsuchiya94,Tsuchiya96}: If the initial conditions is the water-bag form in virial equilibrium, $\Delta$ temporarily decreases and then increases.

\begin{figure}[h]
  \begin{center}
      \includegraphics[scale=0.721]{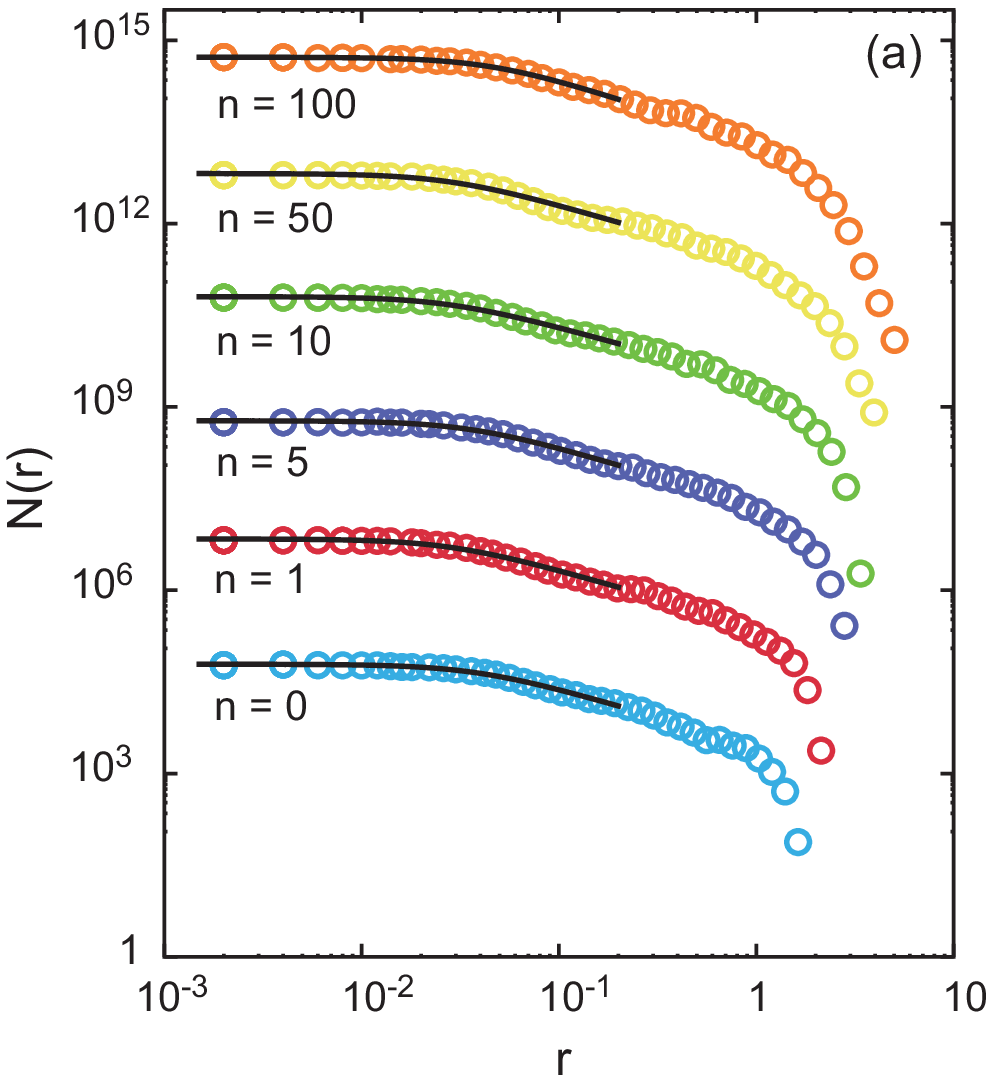} \\
      \includegraphics[scale=0.721]{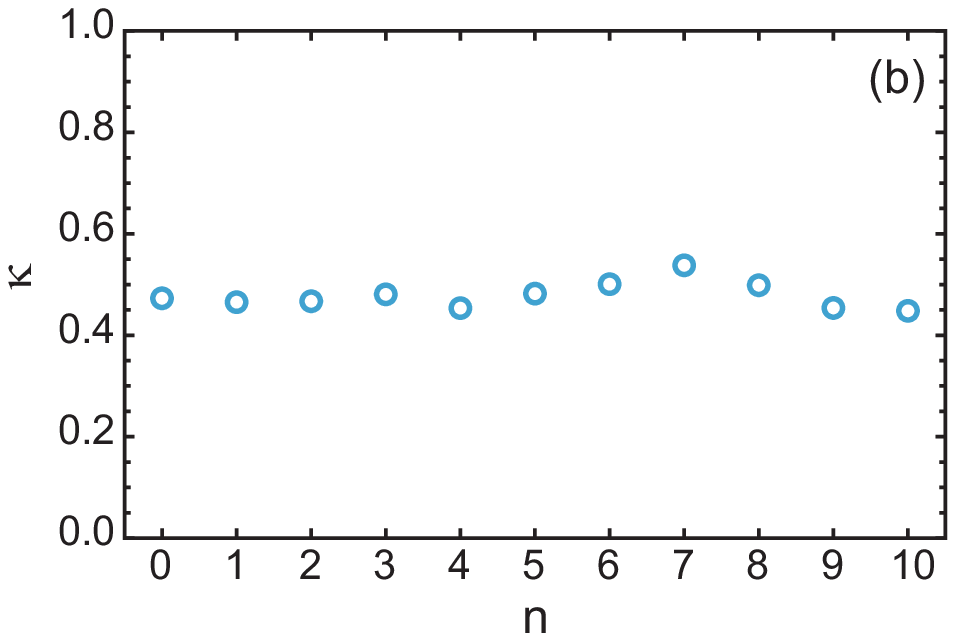} \\
      \includegraphics[scale=0.721]{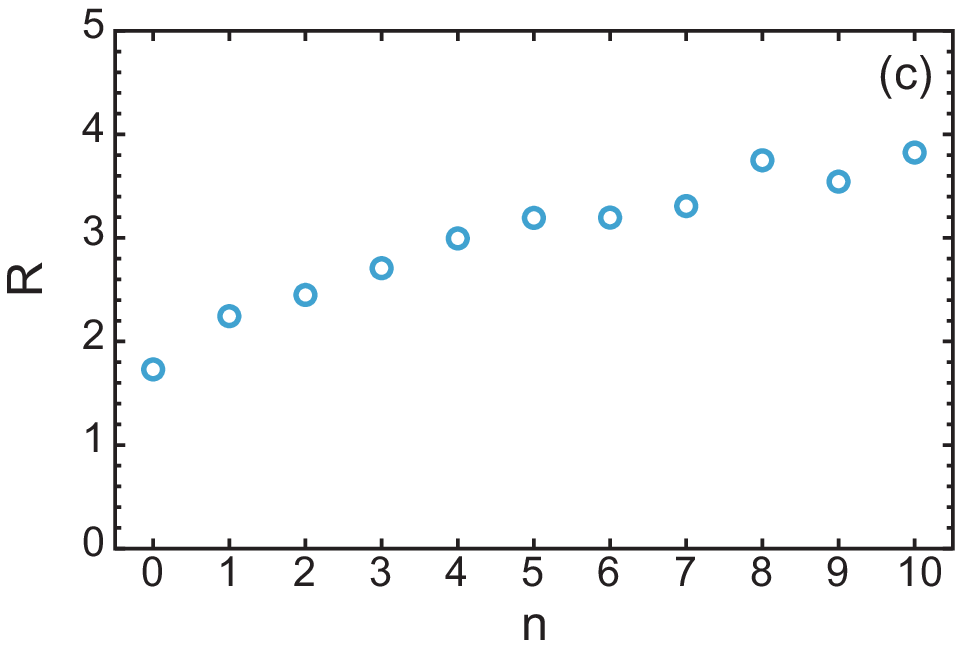} 
    \caption{\label{fig:number}(color online) (a) Number densities of $10^4$ particle systems at qES, for several initial polytrope indices, are plotted by open-circles. The curve passing the circles is a fitting function, Eq.~(\ref{eq:fitf}). 
      Each density is shifted two orders of magnitude up, so that they do not overlay one another. 
(b) The optimum values of ƒÈ in Eq.~(\ref{eq:fitf}) obtained by fitting the number densities with $10^4$ particles at qES with respect to the initial polytrope index $n$.
(c) The radiuses of $10^4$ particle systems at qES with respect to the initial polytrope index $n$.
    }
  \end{center}
\end{figure}

Figure~\ref{fig:number}(a) shows time averaged number densities at qES of $N=10^4$ particles simulations where the null virial condition is satisfied, for several initial polytrope indices ($n=0$, 1, 5, 10, 50 and 100).
So as to be seen easily, each density is shifted by two digits to the below.
Thus, we can find that the number densities around the center of the system  universally have the same form, which can be approximated by Eq.~(\ref{eq:fitf}) shown by black curves in the same figure, independent of the initial polytrope index.
The optimal values of $\kappa$ obtained by fitting the number densities with Eq.~(\ref{eq:fitf}) are plotted in Fig.~\ref{fig:number}(b) as a function of the initial polytrope index $n$.
Since the values are almost 0.5, the expectation stated in Introduction has been verified numerically. 
Contrary to the universality, the radius of the system increases as $n$ gets larger, which can be seen in Figs.~\ref{fig:number}(a) and (c).

\section{envelope equation}

In the previous section, it is found that the number density around the center of the system has the universality depicted by Eq.~(\ref{eq:fitf}) with $\kappa\sim0.5$ when the initial distribution satisfies the null virial condition.
In order to investigate why this condition yields the universality, we utilize the {\it envelope equation} (EEq) of 1DSGS.
If DF, $f(x,u,t)$, of 1DSGS satisfies the Vlasov equation (i.e., in dynamical equilibrium) and it is  origin-symmetric and isotropic in velocity, EEq can be represented as follows\footnote{Note that, although EEq has already been reported for the initial water-bag distribution in Refs.~\cite{Teles11,Levin14}, the derivation in the Letter requires only the symmetry about the origin in phase space.} (see Appendix A):
 \begin{equation}
\ddot{r}_e(t) = \frac{M^2\varepsilon(t)^2}{4r_e(t)^3} - \frac{\Omega(t)}{r_e(t)} \ ,
\label{eq:envelope}
\end{equation}
where $r_e(t) \left(\equiv\sqrt{M\av{x^2}_t}\right)$ is the envelope of the system, $\Omega$ is the total gravitational potential and $\varepsilon(t)^2\equiv4(\av{x^2}_t\av{u^2}_t-{\av{xu}_t}^2)$.
The average $\av{\bullet}_t$ means the integration over phase space with DF: $\av{\bullet}_t \equiv \int \d x\int\d u\bullet f$.
$\varepsilon(t)^2$ cannot be less than 0 because of the Cauchy-Schwarz inequality.

From Eq.~(\ref{eq:envelope}), it is found that the dynamics of the envelope is determined by the opposite effects, collapse and expansion.
If and only if the initial condition of a system satisfies null virial condition (i.e., $\varepsilon(0)=0$), the envelope begins to collapse.
When the system is origin-symmetric in phase space, $\varepsilon(t)^2$ does not change in time:
By using the Vlasov equation, the time derivative of $\varepsilon(t)^2$ is calculated as
$\partial_t{\varepsilon(t)^2} = 8\av{x^2}_t\av{u\partial_x\phi}_t - 8\av{xu}_t\av{x\partial_x\phi}_t$.
Because the system has the symmetry about the origin in phase space, $\int\d uuf = 0$ holds resulting in $\av{u{\partial_x\phi}}_t=\av{xu}_t=0$.
That is, if and only if the null virial condition is satisfied at initial time and the system can be described by the Vlasov equation, the system keeps on collapsing toward the origin.
In other words, all the particles are attracted toward the origin regardless of the initial distribution.

\begin{figure}[h]
  \begin{center}
    \includegraphics[scale=0.851]{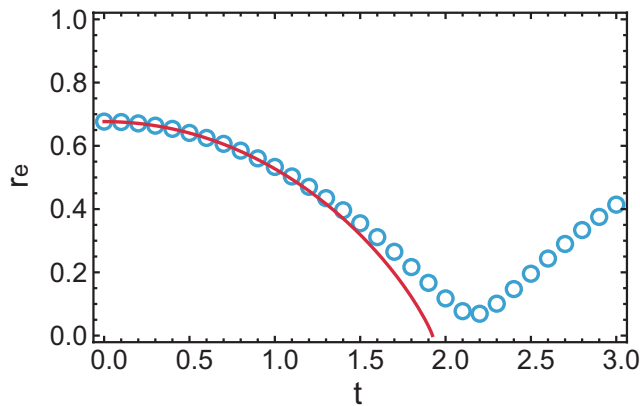}
    \caption{\label{fig:time_evolution_of_envelope}(color online) Time evolution of the envelope of a system, $r_e$. The (blue) open circles are derived from $N=10^4$ particles simulation with $n=1$. The (red) curve means the solution of Eq.~(\ref{eq:envelope}) with setting $\varepsilon^2=0$ and replacing $\Omega(t)$ by $\Omega(0)$.}
  \end{center}
\end{figure}

From Eq.~(\ref{eq:envelope}) with $\varepsilon^2=0$, $r_e$ seems to become 0 within a finite time.
However, the {\it actual} envelope (calculated by the product of $\sqrt{M}$ and the root mean square position which derived from $N$-body simulations) does not reach 0.
The time evolution of the envelope derived from $N=10^4$ particles simulation with $n=1$ is shown in Fig.~\ref{fig:time_evolution_of_envelope} by (blue) open circles.
The solution of Eq.~(\ref{eq:envelope}) with setting $\varepsilon^2=0$ and replacing $\Omega(t)$ by $\Omega(0)$ is also shown in the same figure by the (red) curve.
Although the solution can approximate the behavior of the actual envelope derived from the simulation around $t=0$, it cannot describe the minimum of the actual envelope.

The deviation in behavior of the actual envelope from the solution of Eq.~(\ref{eq:envelope}) with $\varepsilon^2=0$ is due to the increment of $\varepsilon^2$.
The reason why $\varepsilon^2$ becomes non-zero is that it becomes impossible to describe the system by the Vlasov equation, as the system is collapsing toward the origin.
This fact can be understood from the sum of the difference in the one particle energy per unit mass from the initial time which can be calculated as (see Appendix B)
\begin{equation}
\sum_{i=1}^{N}\Delta\mathcal{E}_i(t) = \frac{1}{m}\left\{\Omega(t) - \Omega(0)\right\} \ ,
\end{equation}
where $\Delta\mathcal{E}_i(t)\equiv\mathcal{E}_i(t)-\mathcal{E}_i(0)$ in which $\mathcal{E}_i(t)$ is one particle energy  per unit mass of the $i$th particle at time $t$.
Since the total gravitational potential decreases during the collapse, the sum of the difference in the one particle energy  per unit mass becomes negative.
This means that exchanges of energy occur among the particles, and then, the Vlasov equation cannot describe this situation of the system since the equation is valid for {\it collisionless} SGSs.

\section{Core-Halo structure}
\label{sec:core-halo}
To investigate  the evolution of the system after the collapse caused by the initial null virial condition, let us focus on energy of each particle.

\begin{figure}[h]
  \begin{center}
    \includegraphics[scale=0.851]{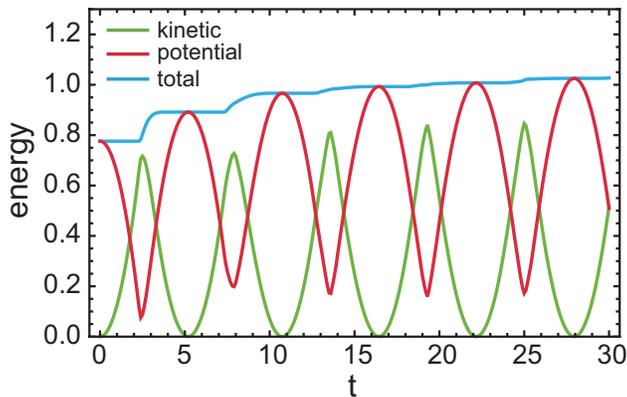}
    \caption{\label{fig:eachene_outer}(color online) Time evolution of the  kinetic, potential and total energy of the outermost at $t=0$ of $N=10^4$ particles simulations with $n=1$.}
  \end{center}
\end{figure}

In Fig.~\ref{fig:eachene_outer}, we show the time evolution of the  kinetic, potential and total energy of the outermost particles at initial time. 
The total energies of particles initially at the outside of the system ({\it outer particles}) gradually increase like this example.

One can simply explain the mechanism of the increment of energy  of the outer particles.
Since the system is origin-symmetric, the gravity acting on a particle at $r$ is proportional to the number of particles within a smaller distance than $r$.
While all the particles move toward the origin owing to the null virial condition, they cannot pass each other.
Note that a difference occurs in the time of arrival at the origin: 
The particles initially at the inside of the system ({\it inner particles}) can reach the origin faster than the outer particles (As discussed in the previous section, the actual envelope does not become 0, which means that all the particles do not reach the origin at the same time).
Therefore, each energy of the outer particles does not change until when the inner particles go through the origin and pass the outer particles.
Then, the outer particles gain the potential energy because of the spread of the inner particles.
After that, the outer particles also spread out from the origin. Because the inner particles return to the origin again, the second passing between the inner and outer particles occurs.
After these movements are repeated for a several times, the system reaches qES.

While the outer particles acquire energy until qES, there are particles losing energy.
This fact can be clear from the following equation,
\begin{equation}
\sum_{i=1}^{N}\Delta\mathcal{E}_i^{\rm qe} = -\frac{1}{3m}\Omega(0) \ ,
\label{eq:energy_balance}
\end{equation}
where $\Delta\mathcal{E}_i^{\rm qe}$ represents the difference of energy of the $i$th particle between the initial time and qES (see Appendix B).
Since the total gravitational potential energy for 1DSGS is positive,  the particles reducing energy must exist.

The movements of the particles gaining and losing energy at qES are as follows.
A region where the lower energy particles can move in real space is restricted to the neighborhood of the origin and we refer to the region as a {\it core}.
Because the particles at the core are influenced by many two-body interactions each other, the trajectories of them change randomly.
Contrary to this, the high energy particles can move in a much wider region in real space which we refer to as a {\it halo}.
Owing to the high energy, the halo particles are not influenced by two-body interaction when passing the core, which means that the movements of them are governed by the mean potential of the system and the trajectories are smooth.
These are a scenario for constructing the core-halo structure of the system satisfying the initial null virial condition.

\section{Fokker-Planck model for the core distribution}
\label{sec:FPM}

Here, we shall derive the core distribution by using a special Fokker-Planck equation.
The Fokker-Planck model approach is appropriate rather than the kinetic equations, i.e., the Boltzmann equation and the Landau equation, because the collision terms of these equations are zero in 1DSGS with equal mass \cite{Eldridge63,Kadomtsev70,Chavanis06,Sano11}.

Before constructing the Fokker-Planck equation, we shall model a force influencing an element of the core particles. That is, we shall begin by constructing a Langevin equation.
Since the system is origin-symmetric, we use $r(t) \equiv |x(t)|$ where $x$ is the position of the element.

Let us introduce the frictional force $-m\gamma \dot{r}(t)$ and the random noise with constant intensity $\sqrt{2D}\xi(t)$, which are essential for a many-body system to reach the thermal equilibrium state, where $\xi(t)$ means a Gaussian-white noise.
Also, the element must be influenced by a mean gravitational force $-F(r)$, which is derived by differentiating $m\phi(r)$ as $-F(r) = -m\partial_r\phi(r)$ where $\phi(r)$ is the mean gravitational potential at qES.
However, this is just a mean gravity. It is natural to consider that the element actually is influenced by a fluctuating gravity around the mean value: The number of particles within a smaller distance than $r$ producing $-F(r)$ through the Poisson equation is the mean value, and the actual number must fluctuate around the value.
This means that another noise which prevents the system from reaching the thermal equilibrium state simply is added to the ordinary Langevin equation, and so this system goes to another stable state, that is, qES.
Therefore, we can regard the noise as distinctive to 1DSGS.

If assuming the intensity of the noise to be a constant denoted by $\sqrt{2\epsilon}$, we can obtain the following Langevin equation:
\begin{equation}
m\ddot{r}(t) + m\gamma\dot{r}(t) = -F(r)\left\{1+\sqrt{2\epsilon}{\eta}(t)\right\} + \sqrt{2D}\xi(t) \ ,
\label{eq:lan1}
\end{equation}
where $\eta$ also means a Gaussian-white noise and has no correlation with $\xi$.  
In the over-damped limit, the equation~(\ref{eq:lan1}) becomes 
\begin{equation}
m\gamma\dot{r}(t)
= -F(r)\left\{1+\sqrt{2\epsilon}{\eta}(t)\right\}
 -\frac{\partial}{\partial r}\frac{\epsilon}{2m\gamma}F(r)^2
+ \sqrt{2D}{\xi}(t) \ .
\end{equation}
The second term on the right hand side of the above equation is a correction term in order to regard products as the Storatonovich product \cite{Sekimoto99}.

The corresponding Fokker-Planck equation is given by
\begin{align}
\frac{\partial}{\partial t}P(r,t) &= \frac{D}{(m\gamma)^2}\frac{\partial^2}{\partial r^2}P(r,t)
 + \frac{1}{m\gamma}\frac{\partial}{\partial r}F(r)P(r,t)+\frac{\epsilon}{(m\gamma)^2}
\frac{\partial^2}{\partial r^2}F(r)^2P(r,t) \ .
\label{eq:FPmodel}
\end{align}
When the system reaches qES, $\partial_tP^{\rm qe}(r)=0$. 
Here, by integrating the Fokker-Planck equation by $r$, we can obtain
\begin{align}
\left\{\frac{D}{(m\gamma)^2}+\frac{\epsilon F(r)^2}{(m\gamma)^2}\right\}
{P^{\rm qe}}'(r)
+\left[\frac{2\epsilon F(r)F'(r)}{(m\gamma)^2} + \frac{F(r)}{m\gamma}\right]P^{\rm qe}(r)
 = \mbox{const.} \ .
\label{eq:st1}
\end{align}

Let us determine the constant of the right-hand side of the above equation by using the boundary condition at $r=0$. 
Because of the symmetry, the mean field force can be represented as $F(r) \propto \int_{0}^{r}\d r'NP^{\rm qe}(r')$. Thus,
$F(0)=0$ where we have used the fact that $P^{\rm qe}(0)$ is bounded.
Moreover, $P^{\rm qe}{}'(0) = 0$, because of the symmetry of the system\footnote{We shall introduce $\tilde{P}^{\rm qe}(x)\d x$ which represents a probability that we can find an element in a range from $x$ to $x+\d x$ at qES. Because of this definition and the symmetry of the system, $\tilde{P}^{\rm qe}(-x) = \tilde{P}^{\rm qe}(x) = P^{\rm qe}(r)/2$. By differentiating the equation with $x (>0) = r$, we can obtain $-\tilde{P}^{\rm qe}{}'(-x) = \tilde{P}^{\rm qe}{}'(x) = {P^{\rm qe}}'(r)/2$. Therefore, $\tilde{P}^{\rm qe}{}'(0) = {P^{\rm qe}}'(0) = 0$.}.
Therefore, the constant is determined to be equal to 0.
In addition, if we use the number density at qES, $\mathcal{N}^{\rm qe} (= NP^{\rm qe})$, the equation~(\ref{eq:st1}) becomes
\begin{align}
{\mathcal{N}^{\rm qe}}'(r)=-\frac{F(r)\left\{m\gamma+2\epsilon F'(r)\right\}}{\left\{{D}+\epsilon F(r)^2\right\}}\mathcal{N}^{\rm qe}(r) \ .
\label{eq:unit1}
\end{align}

The mean gravitational force $F(r)$ satisfies the following equation 
\begin{align}
{F}'(r) =4\pi Gm^2\mathcal{N}^{\rm qe}(r) \ ,
\label{eq:unit2}
\end{align}
which can be derived by substituting the gravitational potential per unit mass $\phi(r)(= \frac{1}{m}\int{\rm d}rF(r))$ into the Poisson equation $\mbox{$\bigtriangleup$}\phi=4\pi G\rho^{\rm qe}=4\pi Gm\mathcal{N}^{\rm qe}(r)$.

Now, we nondimensionalize these equations by using the following units of length and force:
\begin{equation}
{\rm [length]} = \sqrt{3\mathcal{T}/2\pi Gm^3\mathcal{N}^{\rm qe}_0}
\end{equation}
and
\begin{equation}
{\rm [force]} = 2\sqrt{6\pi Gm\mathcal{N}^{\rm qe}_0\mathcal{T}}
\end{equation}
where $\mathcal{T} = D/\{\gamma+56\pi\epsilon Gm\mathcal{N}^{\rm qe}_0\}$ and $\mathcal{N}^{\rm qe}_0\equiv \mathcal{N}^{\rm qe}(0)$.
Then, equations~(\ref{eq:unit1}) and (\ref{eq:unit2}) are altered to
\begin{align}
\bar{\mathcal{N}}^{\rm qe}{}'(\bar{r})=
-\frac{6F(\bar{r})\left\{1+2qF'(\bar{r})\right\}}
{\left\{1+14q+6qF(\bar{r})^2\right\}}\bar{\mathcal{N}}^{\rm qe}(\bar{r}) \ ,
\label{eq:sys1}
\end{align}
and
\begin{equation}
\bar{F}'(\bar{r}) = \bar{\mathcal{N}}^{\rm qe}(\bar{r}) \ ,
\label{eq:sys2}
\end{equation}
where $q\equiv 4\pi\epsilon Gm\mathcal{N}^{\rm qe}_0/\gamma$ and variables with overbar denote dimensionless. We should solve these equations with boundary condition $\bar{\mathcal{N}}^{\rm qe}(0)=1$.

The solution of the equations can be represented as
\begin{align}
\bar{\mathcal{N}}^{\rm qe}(\bar{r})=\frac{1+14q-3f_q^{-1}(\bar{r})^2}
{1+14q+6qf_q^{-1}(\bar{r})^2}
\label{eq:Nsol}
\end{align}
and
\begin{equation}
\bar{F}(\bar{r})=f_q^{-1}(\bar{r})
\label{eq:Fsol}
\end{equation}
where $f_q^{-1}$ is the inverse function of $f_q$ defined as
\begin{equation}
f_q(X) \equiv -2qX+\frac{(1+2q)(1+14q)}{\sqrt{3(1+14q)}}\tanh^{-1}
\left(\sqrt{\frac{3}{1+14q}}X\right) \ .
\label{eq:f}
\end{equation}

\begin{figure}[h]
  \begin{center}
    \includegraphics[scale=0.851]{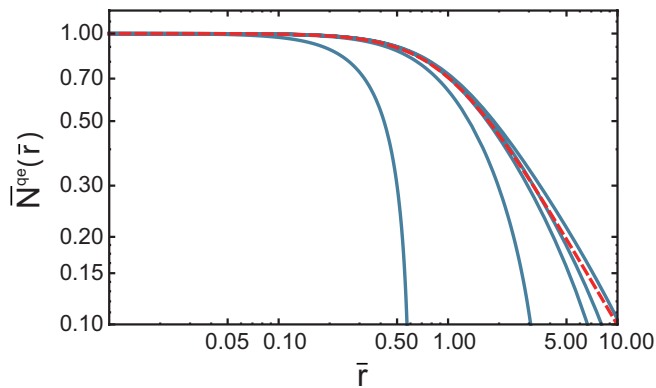}
    \caption{\label{fig:result_mymodel}(color online) Non-dimensional number density Eq.~(\ref{eq:Nsol}) obtained by calculating the inverse function of $f_q$ numerically. Each curve corresponds to the number density with $q = 0$, 1, 3, 4 and 6 respectively from left to right. The (red) dashed curve represents $1/(1+\bar{r}^{2})^{1/2}$.}
  \end{center}
\end{figure}

In Fig.~\ref{fig:result_mymodel}, we show the non-dimensional number density $\bar{\mathcal{N}}^{\rm qe}$, obtained by calculating the inverse function of $f_q$ numerically, as a function of $\bar{r}$.
Each curve corresponds to the number density with $q = 0$, 1, 3, 4 and 6 respectively from left to right. 
The dashed curve, representing $1/(1+\bar{r}^{2})^{1/2}$, means the typical number density around the core at qES found numerically.
Thus, we can describe the numerical results by our model with appropriate $q$.     

We can examine the dependence of $\kappa$ in Eq.~(\ref{eq:fitf}) on $q$ from the analytical formula of the number density at qES.
By approximating the number density around the origin as $\bar{\mathcal{N}}^{\rm qe}(\bar{r})\simeq1/(1+\bar{r}^2)^\kappa$, we can obtain
\begin{equation}
\kappa = \frac{3(1+2q)}{1+14q} \ .
\label{eq:kappa}
\end{equation}
Since $q\ge0$, the range of $\kappa$ is $3/7<\kappa\le3$, which includes the results of numerical simulations.
However, the numerical results show that $\kappa\simeq0.5$ which means that the value of $q$ is limited.
Because $q$ is proportional to a ratio of the intensity of mean gravity fluctuation $\epsilon$ to the friction coefficient $\gamma$, this limited value of $q$ can be regarded as a kind of {\it fluctuation-dissipation relation} \cite{Kubo91}.

Finally, we consider that the number density with $q=0$. Note that $q$ is proportional to $\epsilon$ which prevent the system from reaching the thermal equilibrium state.
Thus, the number density with $q=0$ must agree with the thermal equilibrium state of 1DSGS.

By substituting $q=0$ and Eq.~(\ref{eq:Fsol}) on Eq.~(\ref{eq:f}), we can obtain
\begin{equation}
\bar{r} = \frac{1}{\sqrt{3}}\tanh^{-1}\left[\sqrt{3}\bar{F}(\bar{r})\right] \ .
\end{equation}
Therefore, $\bar{F}$ can be solved as 
\begin{equation}
\bar{F}(\bar{r}) = \frac{1}{\sqrt{3}}\tanh\left[\sqrt{3}\bar{r}\right] \ ,
\end{equation}
and then,
\begin{equation}
\bar{\mathcal{N}}^{\rm qe}(\bar{r}) = 1-3{\bar{F}(\bar{r})}^2 = {\rm sech}^2\left[\sqrt{3}\bar{r}\right] \ ,
\end{equation}
which corresponds to the number density of 1DSGS in the thermal equilibrium state \cite{Rybicki71}.

\section{estimation of the radius of the system}
\label{sec:estimation_R}

In this section, we shall estimate the radius of the system at qES.
As discussed in Sec.~\ref{sec:core-halo}, the particles gaining energy at the initial time construct the halo.
The size of the system can be understood as the range of the movement of the highest-energy particles among them.  
We shall approximate the mechanism about gaining the highest energy as follows: {\it When the outermost particles at initial time reach the origin, the other particles go back to the initial positions instantaneously, and then, the outermost particles gain the highest energy.}
Therefore, the increment of energy of the outermost particles from the initial time to qES can be estimated as
\begin{equation}
\Delta\mathcal{E}_{\rm out}^{\rm qe} = 2\pi Gm\sum_{i=1}^{N}|x_i(0)| \ .
\end{equation}
Since the outermost particles initially have the energy $\mathcal{E}_{\rm out}(0)=2\pi Gm\sum_i|x_i(0)-x_{\rm out}(0)|$ where $x_{\rm out}$ represents the position of the outermost particles, the energy of them at qES can be approximated as
\begin{equation}
\mathcal{E}_{\rm out}^{\rm qe} = \mathcal{E}_{\rm out}(0) + \Delta\mathcal{E}_{\rm out}^{\rm qe} = 2\pi Gm\sum_{i=1}^{N}|x_i(0)-x_{\rm out}(0)| + 2\pi Gm\sum_{i=1}^{N}|x_i(0)| \ .
\end{equation}

The system is assumed to be origin-symmetric.
Then, the energy of the highest-energy particles at rest for a moment becomes $2\pi GM\mathcal{R}_{\rm est}^{\rm qe}$ where $\mathcal{R}_{\rm est}^{\rm qe}$ is the estimated radius of the system.
Hence, the radius of the system can be estimated as
\begin{equation}
  \mathcal{R}_{\rm est}^{\rm qe} = \frac{\mathcal{E}_{\rm out}^{\rm qe}}{2\pi GM} = \frac{m}{M}\sum_{i=1}^{N}\left\{|x_i(0)-x_{\rm out}(0)| + |x_i(0)|\right\} \ .
\label{eq:estimatedR}
\end{equation}

\begin{figure}[h]
  \begin{center}
    \includegraphics[scale=0.851]{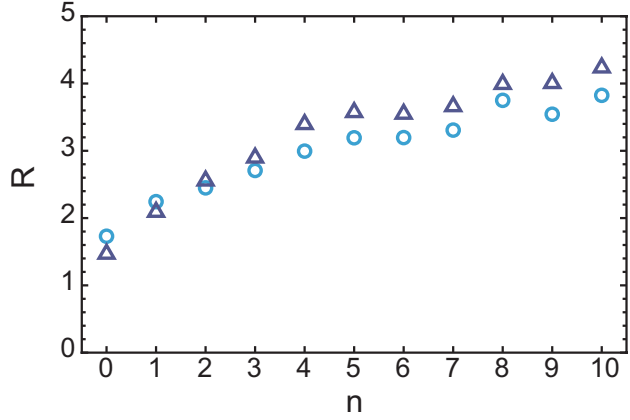}
    \caption{\label{fig:R}(color online) Radiuses of $10^4$ particle systems at qES shown by circles and the estimated radiuses from Eq.~(\ref{eq:estimatedR}) shown by triangles with varying the initial polytrope index $n$.}
  \end{center}
\end{figure}

In Fig.~\ref{fig:R}, the estimated radiuses of $10^4$ particle systems at qES from Eq.~(\ref{eq:estimatedR}) are plotted by triangles with the actual radiuses of the same systems shown by circles with varying the initial polytrope index $n$.
We can see that the estimated radiuses almost corresponds to the actual ones although we roughly approximate the mechanism that the outermost particles gain energy.

\section{description of the whole of density profile at qES by our model}
\label{description}

Now, we shall describe the whole of density profile of 1DSGS at qES by the Fokker-Planck model for the core and the estimated radius as follows.
First, the radius of the system $\mathcal{R}_{\rm est}^{\rm qe}$ is estimated from Eq.~(\ref{eq:estimatedR}) with the initial distribution $x_i(0)$.
Next, we use constrained conditions about the total number and the total energy.
The total number must satisfy 
\begin{equation}
N=a\mathcal{N}^{\rm qe}_0\int_{0}^{\mathcal{R}_{\rm est}^{\rm qe}/a}\d\bar{r}\bar{\mathcal{N}}^{\rm qe}(\bar{r})
\end{equation}
which can be computed as
\begin{equation}
\bar{F}[\mathcal{R}_{\rm est}^{\rm qe}/a] = \frac{N}{a\mathcal{N}^{\rm qe}_{0}} \ ,
\label{eq:relF_N}
\end{equation}
where $a$ represents the unit of length. With this relation and  Eq.~(\ref{eq:f}), the following relation can be obtained:
\begin{equation}
\mathcal{R}_{\rm est}^{\rm qe} = -2\frac{N}{\mathcal{N}^{\rm qe}_{0}}{q} + a
\frac{(1+2q)(1+14q)}{\sqrt{3(1+14q)}}\tanh^{-1}
\left(\sqrt{\frac{3}{1+14q}}\frac{N}{a\mathcal{N}^{\rm qe}_{0}}\right) \ .
\label{eq:constrainedN}
\end{equation}
The total energy $E$ can be expressed only by the total gravitational potential owing to the virial relation as $E=3\Omega^{\rm qe}/2$. Since $\Omega^{\rm qe}$ can be computed by integrating $\bar{\mathcal{N}}^{\rm qe}(\bar{r})$, the constrained condition for $E$ is (see Appendix C)
\begin{equation}
E = \frac{1}{2}\pi GM^2
\left\{
3\mathcal{R}_{\rm est}^{\rm qe} + \frac{2N}{\mathcal{N}^{\rm qe}_{0}}q + \frac{a^2\mathcal{N}^{\rm qe}_{0}}{N}(1+14q)
\left(1-\frac{\mathcal{R}_{\rm est}^{\rm qe}\mathcal{N}^{\rm qe}_{0}}{N}\right)
\right\} \ .
\label{eq:constrainedE}
\end{equation}

Finally, the parameter $q$ the constrained conditions contain must be decided.
Because there are no conditions for deriving the parameter, we shall obtain the value by fitting the power law behavior of the numerical number density at qES with $r^{-2\kappa}$ and utilizing Eq.~(\ref{eq:kappa}). 

\begin{figure}[h]
  \begin{center}
    \includegraphics[scale=0.851]{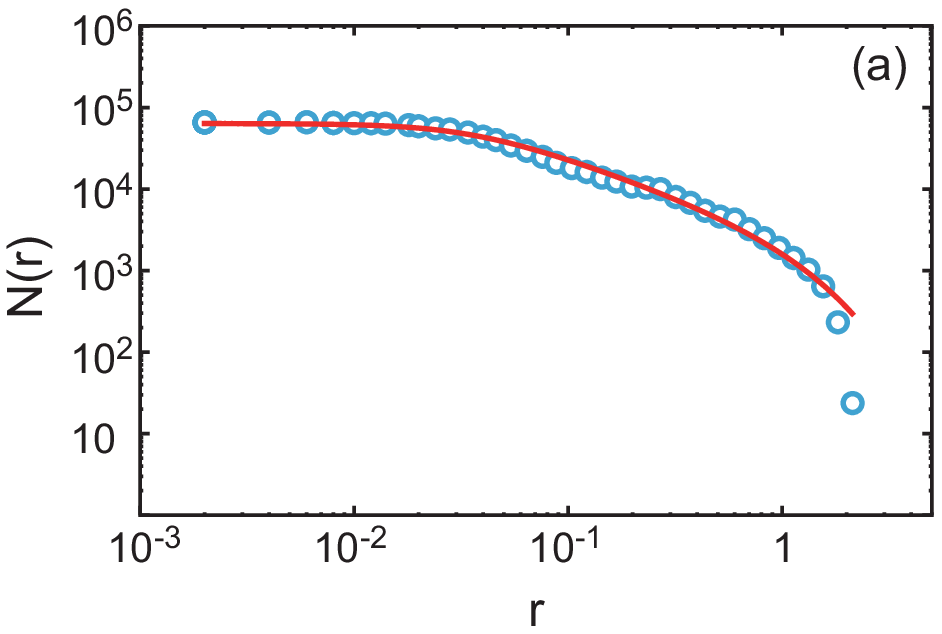} \ 
    \includegraphics[scale=0.851]{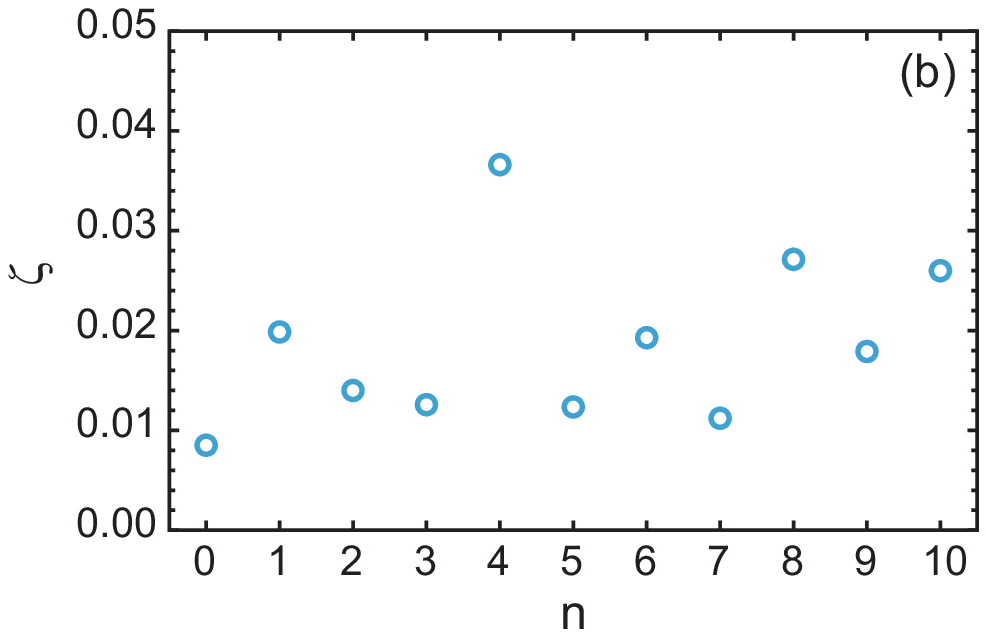}
    \caption{\label{fig:ourmodel}(color online) (a) Number density of $10^4$ particle system with the initial polytope index $n=1$ at qES plotted by open-circles. The (red) curve means the number density of our model derived from the initial distribution of particles, $N$, $E$ and $q$.
(b) Degree of agreement between our model and numerical result defined by Eq.~(\ref{eq:dmodel}) with varying the initial polytrope index $n$.}
  \end{center}
\end{figure}

From the initial distribution of particles, $N$, $E$ and $q$ calculated in this manner, we compute $\mathcal{R}_{\rm est}^{\rm qe}$, $\mathcal{N}^{\rm qe}_{0}$ and $a$ by Eqs.~(\ref{eq:constrainedN}) and (\ref{eq:constrainedE}). 
Figure~\ref{fig:ourmodel}(a) shows the number density of $10^4$ particle system with the initial polytope index $n=1$ at qES by open-circles. The curve means the number density derived by the above method.
We can see that our model can describe $\mathcal{R}_{\rm est}^{\rm qe}$, $\mathcal{N}^{\rm qe}_{0}$ and $a$ well.
Figure~\ref{fig:ourmodel}(b) shows the degree of agreement between our model ($\mathcal{N}^{\rm qe}_0\bar{\mathcal{N}}^{\rm qe}$) and the numerical result ($\mathcal{N}^{\rm num}$) defined by
\begin{equation}
\zeta=\frac{1}{N^2}\int_{0}^{\mathcal{R}_{\rm est}^{\rm qe}}\d r\left\{\mathcal{N}^{\rm qe}_0\bar{\mathcal{N}}^{\rm qe}(r/a)-\mathcal{N}^{\rm num}(r)\right\}^2 \ ,
\label{eq:dmodel}
\end{equation}
with varying the initial polytrope index $n$. 
The same degree of agreement is found for the other initial polytrope indices.

\section{Concluding remarks}
 
In the Letter, we have made it clear numerically that 1DSGS has the qES when the initial conditions satisfy the null virial condition and the density profile around the origin is found to have the universality independent of initial profile in real space.   
The universal density profile around the center is approximated by Eq.~(\ref{eq:fitf}) with $\kappa\simeq0.5$.

The reason why the null virial condition is sufficient for the universality is unveiled by the envelope equation.
In general, this equation contains the opposite effects, collapse and expansion.
If and only if the initial condition  satisfies the null virial condition and the system can be described by the Vlasov equation, the envelope keeps on collapsing toward the origin.
In other words, for the other virial conditions, the envelope is attracted to the other points depending on the initial condition.
After the collapse, the particles exchange energy each other resulting in the core-halo structure: The core and the halo are composed by the particles gaining and losing energy, respectively. 
When the initial distribution is the water-bag form and near virial equilibrium, the exchange of energy also causes the core-halo structure \cite{Teles11}.
Note that, however, a mechanism that the halo particles acquire high energy is a parametric resonance of the oscillating potential and is quite different from what is discussed in the Letter.
    
Since the core particles are influenced by the two-body interactions many times over, the trajectories of them change randomly.
Thus, we have modeled the movement by the special Fokker-Planck equation, Eq.~(\ref{eq:FPmodel}), reflecting a fluctuation of the mean gravitational force. The number density profile derived as a steady solution of Eq.~(\ref{eq:FPmodel}) has no contradiction with the numerical results.

Contrary to this, since the particles of the halo have high energy, the movements are governed by the mean gravitational potential and the trajectories are smooth. We have approximated the mechanism of gaining the highest energy and estimated the radius of the system by the range of the movement of the particles. The estimated radiuses almost correspond to the real ones as shown in Fig.~\ref{fig:R}.

We also have compared the number density profiles of our model with the estimated radius, with the numerical simulation results.
The number density of our model can be obtained by the initial positions of the particles, the total energy, the total number of particles and the index of the power law behavior of the number density.
From Figs.~\ref{fig:ourmodel}(a) and (b), we can see that our model can describe the numerical results well for the case of $n=1$ and the same degree of agreement is found for the other initial polytrope indices.

Although a factor causing $\kappa\sim0.5$ remains unclear, the dynamics to the core-halo stricture is unveiled and the movements at the core and the halo can be explained in the Letter.
Thus, it is expected that the scenario to the universal density profile of 1DSGS will help to clarify the origin of the universality of globular clusters or molecular clouds.
 Moreover, from the fact obtained numerically in the Letter and the universal density profile of 2DSGS and 3DSGS \cite{Tashiro10,Tashiro16}, we can make a conjecture that the density profile of $n$ dimensional SGS at qES from the null virial condition is described by Eq.~(\ref{eq:fitf}) with $\kappa\simeq n/2$.
We can consider that the conjecture gives a new direction to the research about SGS and must be examined.

\section{Acknowledgments}
The author would like to thank members of astrophysics laboratory at Ochanomizu University for extensive discussions.
He also appreciates the advice about parallel computing of $N$-body simulations from T. Takayuki.

\appendix
\section{derivation of the envelope equation}

Since $f(x,u,t)$ satisfies the Vlasov equation, $\partial_tf=-u\partial_xf+\partial_x\phi\partial_uf$, the second derivative of $r_e(t)$ with respect to time can be calculated as
 \begin{equation}
\ddot{r}_e(t) = \frac{M\av{u^2}_t}{r_e(t)} - \frac{M\av{x\partial_x\phi}_t}{r_e(t)}
-\frac{M^2{\av{xu}_t}^2}{r_e(t)^3}
\end{equation}
where the average $\av{\bullet}_t$ means the integration over phase space with DF: $\av{\bullet}_t \equiv \int \d x\int\d u\bullet f$.
By multiplying $M\av{x^2}_t/r_e(t)^2 \ (=1)$ by the first term on R.H.S., the equation becomes
\begin{equation}
\ddot{r}_e(t) = \frac{M^2\varepsilon(t)^2}{4r_e(t)^3} - \frac{M\av{x\partial_x\phi}_t}{r_e(t)} \ ,
\end{equation}
where $\varepsilon(t)^2\equiv 4(\av{x^2}_t\av{u^2}_t-{\av{xu}_t}^2)$.

Let us assume that the system is origin-symmetric in phase space.
Then, with $r\equiv|x|$, $v\equiv|u|$, $\av{x\partial_x\phi}_t$ is calculated as 
\begin{align}
\av{x\frac{\partial\phi}{\partial x}}_t &= \av{r\frac{\partial\phi}{\partial r}}_t \nonumber \\
&= \int_0^{\mathcal{R}}2\d rr\frac{\partial\phi}{\partial r}\int_0^{\mathcal{V}}2\d vf \nonumber \\
&=\frac{1}{2\pi GM}\int_0^{\mathcal{R}}\d rr\frac{\partial\phi}{\partial r}\frac{\partial^2\phi}{\partial r^2}
\end{align}
where $\mathcal{R}$ and $\mathcal{V}$ are respectively the radius and the maximum velocity of the system, and 
we have used the Poisson equation, $\partial_r^2\phi=4\pi G\rho$ in which $\rho=M\int2\d vf$. 

The total gravitational potential, $\Omega \ (=\int2\d r\rho2\pi Gr\int2\d r'\rho)$, also can be calculated as 
\begin{align}
\Omega &=8\pi G\int_0^{\mathcal{R}}\d r\rho(r)r\int_0^{r}\d r'\rho(r') \nonumber \\
&=\frac{1}{2\pi G}\int_0^{\mathcal{R}}\d rr\frac{\partial^2\phi}{\partial r^2}\int_0^{r}\d r'\frac{\partial^2\phi}{\partial r'^2} \nonumber \\
&=\frac{1}{2\pi G}\int_0^{\mathcal{R}}\d rr\frac{\partial\phi}{\partial r}\frac{\partial^2\phi}{\partial r^2} \ ,
\end{align}
where we have used $\left.\partial_r\phi\right|_{r=0}=0$, and then,
\begin{equation}
\frac{\Omega}{M} = \av{x\frac{\partial\phi}{\partial x}}_t \ .
\end{equation}
Therefore, we can obtain EEq,
\begin{equation}
\ddot{r}_e(t) = \frac{M^2\varepsilon(t)^2}{4r_e(t)^3} - \frac{\Omega(t)}{r_e(t)} \ .
\end{equation}

\section{sum of the difference in the one particle energy  per unit mass from the initial time}
The total energy of 1DSGS, $E$,  can be described as 
\begin{equation}
\frac{E}{m} = \sum_{i=1}^{N}\frac{1}{2}u_{i}(t)^2 + \frac{1}{2}\sum_{i,j}2\pi Gm|x_i(t)-x_j(t)|  \ .
\end{equation}
Let us introduce the gravitational potential per unit mass at $x$, $\phi(x)\equiv\sum_{j}2\pi Gm|x-x_j|$, and then, the above equation becomes
\begin{align}
\frac{E}{m} &= \sum_{i=1}^{N}\frac{1}{2}u_{i}(t)^2 + \frac{1}{2}\sum_{i=1}^{N}\phi[x_i(t)] \nonumber \\
 &= \sum_{i=1}^{N}\left\{\frac{1}{2}u_{i}(t)^2+\phi[x_i(t)]\right\} - \frac{1}{2}\sum_{i=1}^{N}\phi[x_i(t)] \nonumber \\
&= \sum_{i=1}^{N}\mathcal{E}_i(t) -\frac{\Omega(t)}{m} \ ,
\end{align}
where $\mathcal{E}_i \equiv \frac{1}{2}{u_i}^2 + \phi(x_i)$.
Thus, the sum of the one particle energy  per unit mass can be depicted as 
\begin{equation}
\sum_{i=1}^{N}\mathcal{E}_i(t) = \frac{1}{m}\left\{E + \Omega(t)\right\} \ .
\end{equation}
Then, the sum of the difference in the one particle energy  per unit mass from the initial stag  can be obtained as
\begin{equation}
\sum_{i=1}^{N}\Delta\mathcal{E}_i(t) = \frac{1}{m}\left\{\Omega(t) - \Omega(0)\right\} \ ,
\label{eq:pre9}
\end{equation}
where $\Delta\mathcal{E}_i(t)\equiv\mathcal{E}_i(t)-\mathcal{E}_i(0)$.

By calculating the difference of this equation between the initial time and qES, we can obtain
\begin{equation}
\sum_{i=1}^{N}\Delta\mathcal{E}_i^{\rm qe} = \frac{1}{m}\left\{\Omega^{\rm qe} - \Omega(0)\right\} \ ,
\label{eq:pre10}
\end{equation}
where $\Delta\mathcal{E}_i^{\rm qe}\equiv\mathcal{E}_i^{\rm qe}-\mathcal{E}_i(0)$ and the superscript qe indicates the quantity at qES.

If the system satisfies the null virial condition at the initial time, the energy conservation law can be written as
\begin{equation}
E = 0 + \Omega(0) = K^{\rm qe} + \Omega^{\rm qe} \ .
\end{equation}
At qES, the virial theorem must hold: $2K^{\rm qe} = \Omega^{\rm qe}$.
Therefore,
\begin{equation}
\Omega^{\rm qe} = \frac{2}{3}\Omega(0) \ ,
\end{equation}
and the equation (\ref{eq:pre10}) becomes
\begin{equation}
\sum_{i=1}^{N}\Delta\mathcal{E}_i^{\rm qe} = -\frac{1}{3m}\Omega(0) \ .
\end{equation}

\section{Constrained condition for the total energy}

The total gravitational potential at qES, $\Omega^{\rm qe}, $ can be computed by integrating $\bar{\mathcal{N}}^{\rm qe}$ as
\begin{align}
\Omega^{\rm qe} &= a^3\int_0^{\bar{\mathcal{R}}^{\rm qe}_{\rm est}}\d\bar{r}m\mathcal{N}^{\rm qe}_0\bar{\mathcal{N}}^{\rm qe}(\bar{r})2\pi G\bar{r}\int_0^{\bar{r}}\d\bar{r}'m\mathcal{N}^{\rm qe}_0\bar{\mathcal{N}}^{\rm qe}(\bar{r}') \nonumber \\
 &= 2\pi G(m\mathcal{N}^{\rm qe}_0)^2a^3\int_0^{\bar{\mathcal{R}}^{\rm qe}_{\rm est}}\d\bar{r}\bar{F}'(\bar{r})\bar{r}\int_0^{\bar{r}}\d\bar{r}'\bar{F}'(\bar{r}') \nonumber \\
&=\pi G(m\mathcal{N}^{\rm qe}_0)^2a^3\left\{\bar{\mathcal{R}}^{\rm qe}_{\rm est}\bar{F}(\bar{\mathcal{R}}^{\rm qe}_{\rm est})^2 - \int_0^{\bar{\mathcal{R}}^{\rm qe}_{\rm est}}\d\bar{r}\bar{F}(\bar{r})^2\right\} \ ,
\end{align}
where $\bar{\mathcal{R}}^{\rm qe}_{\rm est}\equiv \mathcal{R}^{\rm qe}_{\rm est}/a$ and we have used Eq.~(\ref{eq:sys2}) and $\bar{F}(0)=0$.

We can perform the integral of the second term on the last R.H.S of the above equation by substituting $\bar{F}(\bar{r}) = f_q^{-1}(\bar{r}) = X$:
\begin{align}
\int_0^{\bar{\mathcal{R}}^{\rm qe}_{\rm est}}\d\bar{r}\bar{F}(\bar{r})^2&= \int_0^{\bar{F}(\bar{\mathcal{R}}^{\rm qe}_{\rm est})}\d Xf_q'(X)X^2 \nonumber \\
 &= \left[X^2f_q(X)\right]_0^{\bar{F}(\bar{\mathcal{R}}^{\rm qe}_{\rm est})} -2\int_0^{\bar{F}(\bar{\mathcal{R}}^{\rm qe}_{\rm est})}\d Xf_q(X)X \nonumber \\
&= \bar{\mathcal{R}}^{\rm qe}_{\rm est}\bar{F}(\bar{\mathcal{R}}^{\rm qe}_{\rm est})^2 -\frac{1}{3}\bar{F}(\bar{\mathcal{R}}^{\rm qe}_{\rm est})\left\{(1+2q)(1+14q)-4q\bar{F}(\bar{\mathcal{R}}^{\rm qe}_{\rm est})^2\right\} \nonumber \\
&\mbox{}+\frac{\sqrt{3}}{9}(1+2q)\sqrt{1+14q}\left\{1+14q-3\bar{F}(\bar{\mathcal{R}}^{\rm qe}_{\rm est})^2\right\}\tanh^{-1}\left[\sqrt{\frac{3}{1+14q}}\bar{F}(\bar{\mathcal{R}}^{\rm qe}_{\rm est})\right] \nonumber \\
&= \bar{\mathcal{R}}^{\rm qe}_{\rm est}\bar{F}(\bar{\mathcal{R}}^{\rm qe}_{\rm est})^2 -\frac{\bar{F}(\bar{\mathcal{R}}^{\rm qe}_{\rm est})^2}{3}\left[3\bar{\mathcal{R}}^{\rm qe}_{\rm est} + 2q\bar{F}(\bar{\mathcal{R}}^{\rm qe}_{\rm est})+\frac{1+14q}{\bar{F}(\bar{\mathcal{R}}^{\rm qe}_{\rm est})}\left\{1-\frac{\bar{\mathcal{R}}^{\rm qe}_{\rm est}}{\bar{F}(\bar{\mathcal{R}}^{\rm qe}_{\rm est})}\right\}\right] \ ,
\end{align} 
where we have used Eqs.~(\ref{eq:relF_N}) and (\ref{eq:constrainedN}).

Thus, by using Eq.~(\ref{eq:relF_N}) and $E=3\Omega^{\rm qe}/2$, we can obtain
\begin{align} 
\Omega^{\rm qe} &= \frac{1}{3}\pi G(m\mathcal{N}^{\rm qe}_0)^2a^3\bar{F}(\bar{\mathcal{R}}^{\rm qe}_{\rm est})^2\left[3\bar{\mathcal{R}}^{\rm qe}_{\rm est} + 2q\bar{F}(\bar{\mathcal{R}}^{\rm qe}_{\rm est})+\frac{1+14q}{\bar{F}(\bar{\mathcal{R}}^{\rm qe}_{\rm est})}\left\{1-\frac{\bar{\mathcal{R}}^{\rm qe}_{\rm est}}{\bar{F}(\bar{\mathcal{R}}^{\rm qe}_{\rm est})}\right\}\right] \nonumber \\
 &=\frac{1}{3}\pi GM^2
\left\{
3\mathcal{R}_{\rm est}^{\rm qe} + \frac{2N}{\mathcal{N}^{\rm qe}_{0}}q + \frac{a^2\mathcal{N}^{\rm qe}_{0}}{N}(1+14q)
\left(1-\frac{\mathcal{R}_{\rm est}^{\rm qe}\mathcal{N}^{\rm qe}_{0}}{N}\right)
\right\} \ ,
\end{align} 
and Eq.~(\ref{eq:constrainedE}).



\begin{thebibliography}{9}
\bibitem{Tashiro16}
T. Tashiro, {\it Phys. Rev.} E {\bf 93}, 020103(R) (2016).


\bibitem{Tashiro10}
T. Tashiro and T. Tatekawa, {\it J. Phys. Soc. Jpn}, {\bf 79}, pp. 063001, (2010).




\bibitem{Levin14}
Y. Levin, R. Pakter, F. B. Rizzato, T. N. Teles and F. P. C. Benetti, {\it Phys. Rep.} {\bf 535},  1 (2014).


\bibitem{Campa09}
A. Campa, T. Dauxois and  S. Ruffo, {\it Phys. Rep.} {\bf 480}, 57 (2009).

\bibitem{Arzoumanian11}
D. Arzoumanian et. al., {\it A{\rm \&}A}, {\bf 529}, L6 (2011).

\bibitem{Stepnik03}
B. Stepnik et. al., {\it A{\rm \&}A}, {\bf 398}, 551 (2003).


\bibitem{Peterson75}
C. J. Peterson and I. R. King, {\it Astron. J.} {\bf 80}, 427 (1975). 
\bibitem{Trager95}
S. C. Trager, I. R. King, and S. Djorgovski, {\it Astron. J.} {\bf 109}, 218 (1995). 

\bibitem{Binney08}
J. Binney and S. Tremaine, {\it Galactic Dynamics'}: (Second Edition), Princeton University Press, Princeton, NJ, 2008.



 
\bibitem{Rybicki71}
  G. B. Rybicki, {\it Astrophys. Space Sci.} {\bf 14} 56 (1971).

 
\bibitem{Hohl67}
  F. Hohl and D. Tilghman Broaddus, {\it Phys. Lett.} A {\bf 25} 713 (1967). 

\bibitem{Hohl67b}
F. Hohl and M. R. Feix, {\it Astrophys.} J. {\bf 147} 1164 (1967).

\bibitem{Hohl68}
F. Hohl and J. W. Campbell, {\it Astron.} J. {\bf 73} 611 (1968).



\bibitem{Wright82}
    H. L. Wright, B. N. Miller, and W. E. Stein, {\it Astrophys. Space Sci.} {\bf 84} 421 (1982).


\bibitem{Luwel85}
  M. Luwel and G. Severne, {\it Astron. Astrophys.} {\bf 152}  305 (1985). 


\bibitem{Luwel84}
  M. Luwel, G. Severne, and P. J. Rousseeuw, {\it Astrophys. Space Sci.} {\bf 100}  261 (1984).

\bibitem{Severne84}
  G. Severne, M. Luwel, and P. J. Rousseeuw, {\it Astron. Adsrophys.} {\bf 138}  365 (1984).

\bibitem{Reidl91}
  C. J. Reidl Jr. and B. N. Miller, {\it Astrophys. J.}  {\bf 371} 260 (1991).
  

\bibitem{Yamashiro92}
  T. Yamashiro, N. Gouda and M. Sakagami, {\it Progress of Theoretical Physics} {\bf 88} 269 (1992).

\bibitem{Tsuchiya94}
    T. Tsuchiya, T. Konishi, and N. Gouda, {\it Phys. Rev.} E {\bf 50} 2607 (1994).

\bibitem{Tsuchiya96}
    T. Tsuchiya, T. Konishi, and N. Gouda, {\it Phys. Rev.} E {\bf 53} 2210 (1996).


  \bibitem{Lecar71}
    M. Lecar and L. Cohen, {\it Astrophys. Space Sci.} {\bf 13} 397 (1971).
    
\bibitem{Yamaguchi08}
    Y. Y. Yamaguchi, {\it Phys. Rev.} E {\bf 78} 041114 (2008).

\bibitem{Joyce11}
M. Joyce and T. Worrakitpoonpon, {\it Phys. Rev.} E {\bf 84} 011139 (2011).


\bibitem{deBuyl11}
P. de Buyl, D. Mukamel, and S. Ruffo, {\it Phys. Rev.} E {\bf 84} 061151 (2011).

\bibitem{Ribeiro-Teixeira14}
A. C. Ribeiro-Teixeira, et. al., {\it Phys. Rev.} E {\bf 89} 022130 (2014).

\bibitem{Benetti14}
F. P. C. Benetti, et. al., {\it Phys. Rev. Lett.} {\bf 113} 100602 (2014).



\bibitem{Teles11}
    T. N. Teles, Y. Levin and R. Pakter, {\it Mon. Not. R. Astron. Soc.} {\bf 417} L21 (2011).


\bibitem{Eldridge63}
O. C. Eldridge and M. Feix, {\it Phys. Fluids} {\bf 6} 398 (1963).

\bibitem{Kadomtsev70}
B. B. Kadomtsev and O. P. Pogutse, {\it Phys. Rev. Lett.} {\bf 25} 1155 (1970).

\bibitem{Chavanis06}
P. H. Chavanis, {\it Physica A} {\bf 361} 81  (2006).

\bibitem{Sano11}
M. M. Sano and K. Kitahara, {\it J. Phys. Soc. Jpn.} {\bf 80} 084001 (2011) .


\bibitem{Tarcisio10}
T.  N.  Teles, Y.  Levin, R. Pakter and F. B. Rizzato, {\it J. Stat. Mech.}, {\bf 2010}, pp. P05007 (2010).



\bibitem{Sekimoto99}
K. Sekimoto, {\it J. Phys. Soc. Jpn}, {\bf 68}, pp. 1448 (1999).


\bibitem{Kubo91}
R. Kubo, M. Toda and N. Hashitsume, {\it Statistical Physics II: Nonequilibrium Statistical Mechanics},
(Springer-Verlag, Berlin, 1991).

\end{thebibliography}
\end{document}